\documentclass[journal]{IEEEtran}
\usepackage{cite}

\usepackage{mhchem} 

\usepackage[dvips]{graphicx}
\usepackage{color}
\usepackage{amsmath}
\usepackage{stfloats}
\fnbelowfloat
\begin{document}
	
\title{Field-Effect Transistor Based on MoSi$_2$N$_4$ and WSi$_2$N$_4$ Monolayers Under Biaxial Strain: A Computational Study of the Electronic Properties}

\author{Nayereh Ghobadi, Manouchehr Hosseini, and Shoeib Babaee Touski
\thanks{N. Ghobadi is with the Department of Electrical Engineering, University of Zanjan, Zanjan, Iran. M. Hosseini is with the Department of Electrical Engineering, Bu-Ali Sina University, Hamedan, Iran e-mail:(m.hosseini@basu.ac.ir). S. B. Touski is with the Department of Electrical Engineering, Hamedan University of Technology, Hamedan, Iran. }% <-this % stops a space
}

\maketitle

\begin{abstract}
The electronic properties of a field-effect transistor with two different structures of MoSi$_2$N$_4$ and WSi$_2$N$_4$ monolayers as the channel material in the presence of biaxial strain are investigated. The band structures show that these compounds are semiconductors with an indirect bandgap. Their band gaps can be adjusted by applying in-plane biaxial strain. In the following, the variation of the energies of the valleys and corresponding effective masses with respect to the strain are explored. Finally, the strained MoSi$_2$N$_4$ or WSi$_2$N$_4$ are used as the channel of a p-type FET and the corresponding current-voltage characteristic is explored. The results show this FET has an I$\mathrm{_{ON}}$/I$\mathrm{_{OFF}}$ ratio larger than $\mathrm{10^6}$ and subthreshold swing in the range of 96-98 mV/dec. The I$\mathrm{_{ON}}$/I$\mathrm{_{OFF}}$ ratio of these compounds with respect to strain are compared.
\end{abstract}

\begin{IEEEkeywords}
MoSi$_2$N$_4$, WSi$_2$N$_4$, Biaxial Strain, Field-Effect Transistor, I$\mathrm{_{ON}}$/I$\mathrm{_{OFF}}$ ratio, Sub-threshold swing.
\end{IEEEkeywords}

\IEEEpeerreviewmaketitle

\section{Introduction}
\IEEEPARstart{T}{wo-}dimensional materials are being widely studied from 2004 with the successful experimental isolation of graphene \cite{novoselov2004electric}. Graphene with full-sp2 carbon atoms exhibits the remarkable electronic properties \cite{lin2010100,wang2009graphene,dean2010boron}, however, absence of an energy gap seriously precluded its exploitation in electronic applications \cite{zhang2009direct,gava2009ab}. Efforts to create an energy gap in graphene were ineffective \cite{zhang2009direct,son2006energy,giovannetti2007substrate}, and researches extend into other two-dimensional materials such as the transition metal dichalcogenides (TMDs) \cite{fu20212d,hosseini2015strain,hosseini2019investigation}, indium selenide \cite{guo2017band}, indium telluride \cite{touski2020electrical}, and phosphorene \cite{elahi2015modulation} with appropriate energy gaps. The design of new 2D materials with attractive piezoelectricity and flexoelectricity is very attractive to expand the practical application of two-dimensional materials \cite{ghosh2020piezoelectricity,springolo2020flexoelectricity}.

\begin{figure}
	\centering
	\includegraphics[width=0.8\linewidth]{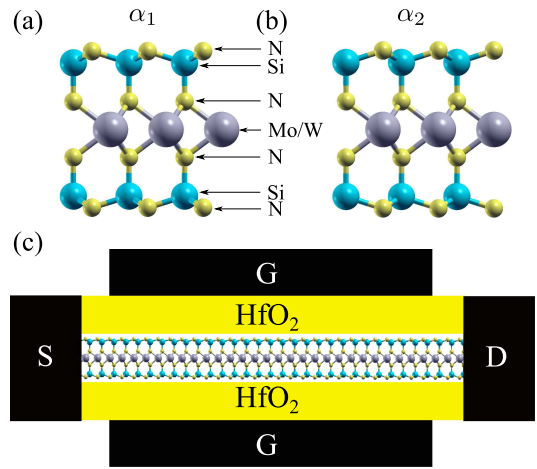}
	\caption{Schematic of MoSi$_2$N$_4$ and WSi$_2$N$_4$ with (a) $\alpha_1$- and (b) $\alpha_2$-configurations. (c) The configuration of the studied DGFET with MoSi$_2$N$_4$ and WSi$_2$N$_4$ as the channel. }
	\label{fig:schematic}
\end{figure}

Single-layer MA$_2$Z$_4$ (M = early transition metal, e.g. Mo, W, and Nb; A = Si or Ge, Z = N, P or As), is a family of materials with covering semiconducting, metallic and magnetic properties \cite{wang2020designing,wang2021,zhong2021strain}. A single layer MoA$_2$Z$_4$ is composed of an M-A sub-layer sandwiched between two A-Z sub-layers. Researchers are predicted high lattice thermal conductivity of 440 and 500 W/mK \cite{mortazavi2021exceptional}, and elastic modulus of 487 and 506 GPa for a single layer of MoSi$_2$N$_4$ and WSi$_2$N$_4$ which have made them attractive 2D materials.
Synthesizing with a large size of 2D MoSi$_2$N$_4$ \cite{hong2020chemical}, the excellent ambient stability of it \cite{hong2020chemical,cai2014polarity}, besides the large theoretical electron/hole mobilities (up to ~ 270/1200 cm$^2$/(Vs) which are near to six times larger than those of monolayer MoS$_2$) creates a bright and versatile future for this material \cite{hong2020chemical,cai2014polarity}. MoSi$_2$N$_4$ and WSi$_2$N$_4$ have an indirect bandgap with the values of 1.73 eV and 2.06 eV, respectively \cite{wang2020designing,touski2021vertical}. This range of bandgap makes them promising candidates for potential optical applications in the visible range. This family has a pair of valley pseudospins. Therefore, these materials have suitable valleytronic properties to be applied in multiple information processing in the future \cite{yang2021valley}. 

It is expected that MA$_2$Z$_4$ family demonstrate potential applications ranging from electronics, optoelectronics, photonics, spintronics, catalysis to energy harvesting\cite{mortazavi2021exceptional}.
With sandwiching a TMD-type MZ$_2$ monolayer into InSe-type A$_2$Z$_2$ in MA$_2$Z$_4$ monolayers, twelve different structures have been achieved \cite{wang2021}. It has been reported that $\alpha_1$ and $\alpha_2$ constructions of MoSi$_2$N$_4$ and WSi$_2$N$_4$ are the most stable structures \cite{wang2021}. It has been shown that this material can withstand strain values up to 19.5 percentage\cite{li2021strain}. The effect of biaxial strain on the electronic properties of these stable constructions as the channel material of a FET has not been explored. Therefore, in this work, we have investigated the electronic properties of $\alpha_1$ and $\alpha_2$ constructions of MoSi$_2$N$_4$ and WSi$_2$N$_4$ in the presence of in-plane biaxial strain with first-principles calculations. In the following, we use a ballistic analytical model as the top-of-the-barrier model to assess the DC performance of the double-gate field-effect transistor (DGFET) based on single-layer MoSi$_2$N$_4$ and WSi$_2$N$_4$ under biaxial strain \cite{rahman2003theory,hosseini2018strain}.

\begin{figure*}
	\centering
	\includegraphics[width=0.8\linewidth]{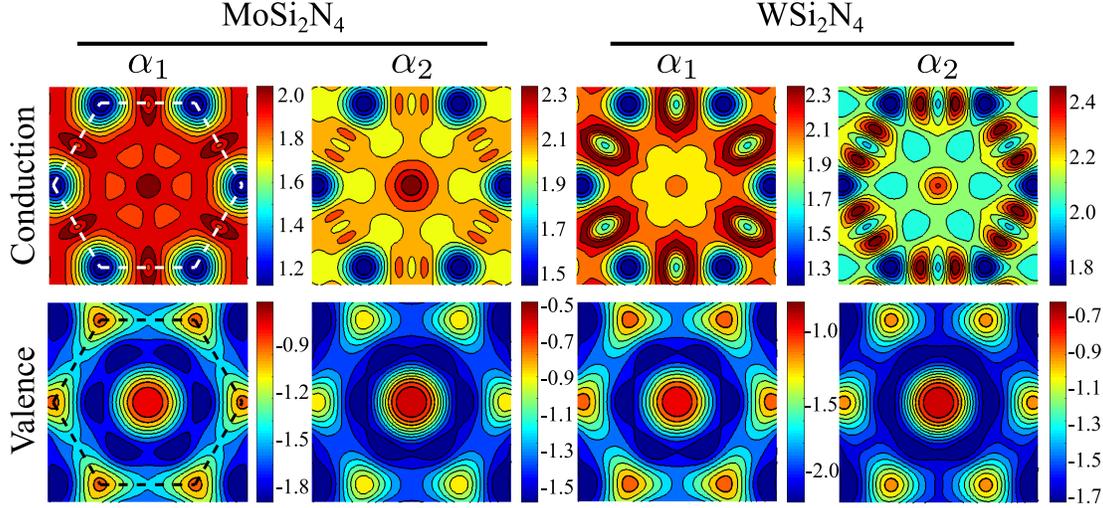}
	\caption{The two-dimensional energy map of the conduction and valence bands for all compounds. The first brilloiun zone is indicated for both conduction and valence band of $\alpha_1$-MoSi$_2$N$_4$. }
	\label{fig:con-val}
\end{figure*}

Some details on the ab initio calculations of the electronic properties in the presence of strain are discussed in Section II. In Section III, The effects of biaxial strain on the electronic properties have been explored and DC performance of the DGFETs based on these materials have been investigated in section IV. Finally, concluding remarks are presented in Section VI.

\section{Computational details}
In order to investigate the properties of MoSi$_2$N$_4$ and WSi$_2$N$_4$ monolayers, density functional calculations are performed using the SIESTA software \cite{soler2002siesta}. The generalized gradient approximation (GGA) with the Perdew-Burke-Ernzerhof (PBE) \cite{perdew1981self} functional is employed for the exchange-correlation term. A Monkhorst-Pack k-point grid of $21\times21\times1$ is chosen for the unit cell. A double-$\zeta$ plus polarization basis-set is used and the energy cutoff is 200 Ry. The total energy is converged to better than $10^{-5}$ eV. The structures are fully relaxed until the force on each atom is less than 0.02 eV/$\mathrm{\AA}$.  To avoid interactions in the normal direction, a vacuum region of 30 $\mathrm{\AA}$  is added. To visualize the atomic structures, the XCrySDen package has been used \cite{kokalj2003computer}. The in-plain biaxial strain is defined as $\varepsilon=(a-a_0)/a_0$, where $a_0$ and $a$ are the equilibrium and deformed lattice constants, respectively. The effective mass of the carriers is calculated using the following equation \cite{touski2020interplay},
\begin{equation}
m^*=\hbar^2/\left(\partial^2E/\partial k^2\right)
\end{equation}
Here, $\hbar$ is the reduced Planck constant, E and k are the energy and wave vector of conduction band minimum and valence band maximum.

\begin{table}[t]
	\caption{ The lattice constant (a), the M-N (d$_{M-N}$) and Si-N (d$_{Si-N}$) bond lengths, the vertical distance between Si and N atoms ($\Delta_{Si-N}$), the thickness and the cohesive energies (E$_{coh}$) of MoSi$_2$N$_4$ and WSi$_2$N$_4$ monolayers. \label{tab:tab1}}
	
	\begin{tabular}{p{1.6cm}p{0.8cm}p{1.0cm}p{0.9cm}p{0.9cm}p{1.2cm}}
		\hline
		\hline
		&a
		($\mathrm{\AA}$) & d$_{M-N}$
		($\mathrm{\AA}$)  & d$_{Si-N}$
		($\mathrm{\AA}$) & $\Delta_{Si-N}$
		($\mathrm{\AA}$) & E$_{coh}$
		(eV/atom)  \\
		\hline
		$\alpha_1$-MoSi$_2$N$_4$ &2.935 &2.122 &1.764 &0.53 &-8.855   \\
		
		\hline
		$\alpha_2$-MoSi$_2$N$_4$ &2.926 &2.122 &1.774 &0.545 &-8.834  \\
		
		\hline
		$\alpha_1$-WSi$_2$N$_4$ &2.949 &2.137 &1.761 &0.519 & -8.942 \\
		
		\hline
		$\alpha_2$-WSi$_2$N$_4$ &2.939 &2.141 &1.773 &0.537 & -8.919    \\

		\hline
		\hline
	\end{tabular}
\end{table}

\section{Results and discussion}

\begin{figure*}
	\centering
	\includegraphics[width=1.0\linewidth]{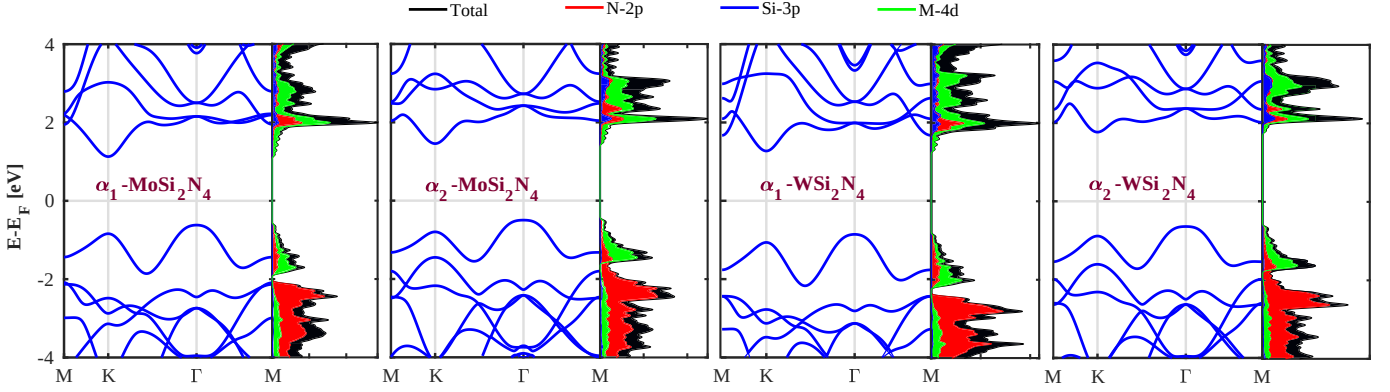}
	\caption{The band structures of MoSi$_2$N$_4$ and WSi$_2$N$_4$ with $\alpha_1$- and $\alpha_2$-configuration. The projected density of states are plotted alongside of its corresponding band structure.  }
	\label{fig:bands}
\end{figure*}

%
%\begin{figure}
%	\centering
%	\includegraphics[width=0.7\linewidth]{bandgap}
%	\caption{The band gap variations as a function of in-plane biaxial strain for four compounds.}
%	\label{fig:bandgap}
%\end{figure}

Two different configurations of MoSi$_2$N$_4$ and WSi$_2$N$_4$ have been reported to have the lowest energy \cite{wang2021}. However, the energy difference between these two structures is small. The schematic of MoSi$_2$N$_4$ and WSi$_2$N$_4$ with these two constructions, $\alpha_1$ and $\alpha_2$, have been shown in Fig. \ref{fig:schematic}. A MoN$_2$ sub-layer is sandwiched between two SiN sub-layers. In the $\alpha_2$-structure, these three sub-layers are exactly placed over others, whereas two SiN sub-layers have shift respect to MoN$_2$ sub-layer in $\alpha_1$-structure. 
The configuration of the studied FET is sketched in Fig. \ref{fig:schematic}(c). A monolayer MoSi$_2$N$_4$ or WSi$_2$N$_4$ and a high-K thin-film HfO$_2$ have been used as the channel material and gate dielectric, respectively. 

The structural properties of these four materials are reported in Table \ref{tab:tab1}. The lattice constants of these four materials are in the same range and are compatible with previously reported works\cite{hong2020chemical,wang2021}. However, $\alpha_2$-structure demonstrates a slightly lower lattice constant than $\alpha_1$ one. On the other hand, the cohesive energies are also in the same range, whereas $\alpha_1$ demonstrates lower cohesive energy and is more stable. In addition, WSi$_2$N$_4$ has a lower cohesive energy than MoSi$_2$N$_4$. The thickness of these compounds is approximately the same and $\alpha_2$-configuration has a higher thickness. Finally, the buckling height of the SiN sub-layer demonstrates a higher value in $\alpha_2$-configuration.

Two-dimensional map of conduction and valence bands in the first Brillouin zone for all compounds is plotted in Fig. \ref{fig:con-val}. It is obvious that the conduction band minimum (CBM) is located at K-valley for all these materials. However, WSi$_2$N$_4$ displays a local minimum at M-valley with an energy close to K-valley. On the other hand, the valence band maximum (VBM) is located at $\Gamma$-valley, whereas the energy of K-valley is close to $\Gamma$-valley and contributes to the VBM. The contours around $\Gamma$-valley are isotropic and circular, whereas K-valley demonstrates anisotropic behavior and is close to triangular.

The band structures along the high symmetry points for four compounds are depicted in Fig. \ref{fig:bands}. The CBM is located at K-valley in all compounds, whereas in $\alpha_2$-structures, the energy of M-point is close to CBM, especially for $\alpha_2$-WSi$_2$N$_4$. On the other hand, the VBM is placed at $\Gamma$-point and the energy of K-point is close to VBM. The corresponding projected density of states (PDOS) are plotted beside the band structures. PDOS demonstrates that both conduction and valence bands are highly comprised of the d-orbitals from Mo or W elements. In addition, the p-orbitals of N atoms also contribute to the conduction and valence bands. The p-orbitals of Si atoms have an influence on the higher energies in the conduction band and do not contribute to the CBM and VBM. 

\begin{table*}[t]
	\caption{The calculated band gaps (E$_g$), position of VBM and CBM, the energies of conduction (E$_C$-E$_F$) and valence band (E$_V$-E$_F$) edges, and the effective masses at K and $\Gamma$-point of the valence band and K-valley of the conduction band. \label{tab:tab2}}
	\begin{tabular}{p{1.6cm}p{0.8cm}p{0.6cm}p{0.6cm}p{1.1cm}p{1.1cm}p{1.2cm}p{1.2cm}p{1.2cm}p{1.2cm}p{1.2cm}p{1.2cm}}
		\hline
		\hline
		& E$_g$ &  VBM & CBM & E$_C$-E$_F$ & E$_V$-E$_F$ &$m^{v,*}_{K\rightarrow M}$ & $m^{v,*}_{K\rightarrow \Gamma}$ & $m^{v,*}_{\Gamma\rightarrow K}$ & $m^{v,*}_{\Gamma\rightarrow M}$&$m^{c,*}_{K\rightarrow M}$ & $m^{c,*}_{K\rightarrow \Gamma}$      \\
		& (eV) &  &  & (eV) & (eV)  &($m_0$)&($m_0$)&($m_0$)&($m_0$)&($m_0$)&($m_0$)  \\
		\hline

		$\alpha_1$-MoSi$_2$N$_4$	&1.748 &$\Gamma$ & K& 1.129&-0.619 &0.838 &0.621 &1.319 &1.314&0.626&0.551 \\
		
		\hline
		
		$\alpha_2$-MoSi$_2$N$_4$	&1.954 &$\Gamma$ &K &1.456 &-0.497 &0.944	 &0.814 &1.589 &1.61&0.639&0.628 \\
		
		\hline
		
		$\alpha_1$-WSi$_2$N$_4$	&2.129 &$\Gamma$ &K &1.272 &-0.857 &0.728 &0.569 &1.206 &1.196&0.525&0.519  \\
		
		\hline
		
		$\alpha_2$-WSi$_2$N$_4$	&2.41 &$\Gamma$ &K &1.766 &-0.644 &0.797 &0.727 &1.52 &1.545&0.578&0.753 \\

		\hline
		\hline
	\end{tabular}
\end{table*}

The electrical properties of these four monolayers are listed in Table \ref{tab:tab2}. The bandgap of $\alpha_1$-configuration of MoSi$_2$N$_4$ is 1.748 eV that is in good agreement with previous results \cite{hong2020chemical,wang2021}. However, $\alpha_2$ one has a higher bandgap as 1.954 eV that is approximately 0.2 eV higher than $\alpha_1$. The bandgap of $\alpha_1$-WSi$_2$N$_4$ also is 2.129 eV that is close to the previous reported values \cite{hong2020chemical,wang2021}. The bandgap increases to 2.41 eV for $\alpha_2$-structures that is 0.3 eV larger than its $\alpha_1$-configuration. The results prove that $\alpha_2$ constructions demonstrate a higher bandgap. As one can see, the CBM and VBM are located at K- and $\Gamma$-valleys, respectively. The effective masses of important valleys in the conduction and valence bands are reported in Table \ref{tab:tab2}. In the valence band, the effective masses of $\Gamma$-valley in different directions are approximately the same, whereas K-valley demonstrates various effective masses along different directions. Previously, we have observed in Fig. \ref{fig:con-val} isotropic and anisotropic contours around $\Gamma$- and K-point, respectively. $\Gamma$-valley demonstrates a higher effective mass with respect to K-valley that is compatible with the corresponding band structure. In addition, $\alpha_2$-structures show larger effective masses along with its higher bandgap. For example, the effective mass of $\alpha_1$-MoSi$_2$N$_4$ in the $\Gamma$-valley is 1.31 m$_0$, while increases to 1.6 m$_0$ for $\alpha_2$ one. Although $\alpha_2$-structure contains the higher effective mass, this configuration shows lower anisotropicity at K-valley. For instance, the effective masses of $\alpha_1$-WSi$_2$N$_4$ are 0.728 and 0.569 m$_0$ and for $\alpha_2$-WSi$_2$N$_4$ are 0.797 and 0.727 m$_0$ along K$\rightarrow$M and K$\rightarrow\Gamma$ paths of valence band, respectively. The difference between the effective masses along different paths for $\alpha_1$- and $\alpha_2$-structures are 0.16 and 0.07 m$_0$ that proves $\alpha_2$-structure demonstrates lower anisotropicity at K-valley.

\begin{figure*}
	\centering
	\includegraphics[width=1.0\linewidth]{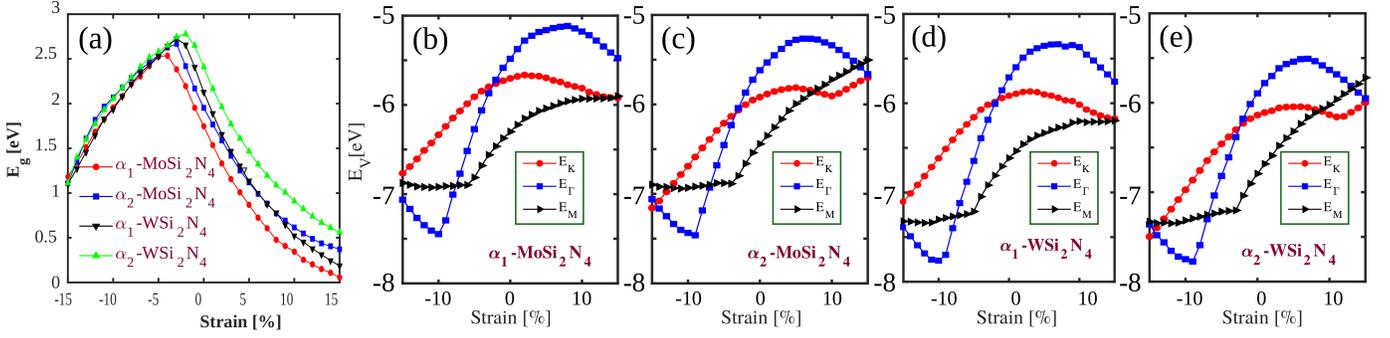}
	\caption{ (a) The band gap variations as a function of in-plane biaxial strain for four compounds. (b-e) The energies of the valence band valleys as a function of the biaxial strain for $\Gamma$-valley ($\mathrm{E_V}$), K-valley ($\mathrm{E_K}$), and M-valley ($\mathrm{E_M}$). }
	\label{fig:energy}
\end{figure*}

The biaxial in-plane strain has been known as a powerful tool to modify the electrical and electronic properties of 2D materials \cite{touski2020interplay,touski2021structural}. The biaxial strain has been applied to four compounds and their electronic properties are explored. First of all, the band gaps of these four materials with respect to the strain are presented in Fig. \ref{fig:energy}(a). The band gaps of all compounds show a maximum at a compressive strain between -2$\%$ and -4$\%$. Then, the band gaps decrease under compressive and tensile strains. The strain value is swept from -15 to 15$\%$. While the bandgap of $\alpha_1$-MoSi$_2$N$_4$ closes under a tensile strain of 15$\%$, the bandgap is higher than 1eV at a compressive strain of 15$\%$. Under compressive strains lower than -5$\%$, the band gaps of all compounds are approximately the same, whereas the band gaps exhibit different values on the other side of the maximum point. The band gaps also decrease more rapidly at low tensile strains and the $\alpha_1$-structures show the lowest band gaps in large tensile strains.

It has been reported in the experimental works that monolayer MoSi$_2$N$_4$ demonstrates a p-type characteristic \cite{hong2020chemical}. In this regard, the valence band (hole) properties are investigated in the following. For clarifying the behavior of the valleys, the energies of the effective valleys in the valence bands with respect to strain are plotted in Fig. \ref{fig:energy}(b-e). Three valleys ($\Gamma$, K and M) participate at the valence band. The VBM is located at $\Gamma$-valley in the tensile strain, whereas K- and M-valleys also contribute to the VBM of $\alpha_2$-structures at the large tensile strains. At these strains, the M-valley is dominant and two other valleys also contribute VBM. $\Gamma$- and K-valleys on valence band possess the close energy at small compressive strains while the K-valley is dominant at compressive strains. Similar to the large tensile strains, three valleys contribute to the VBM of $\alpha_2$-structures at the large compressive strains. 

\begin{figure}
	\centering
	\includegraphics[width=0.8\linewidth]{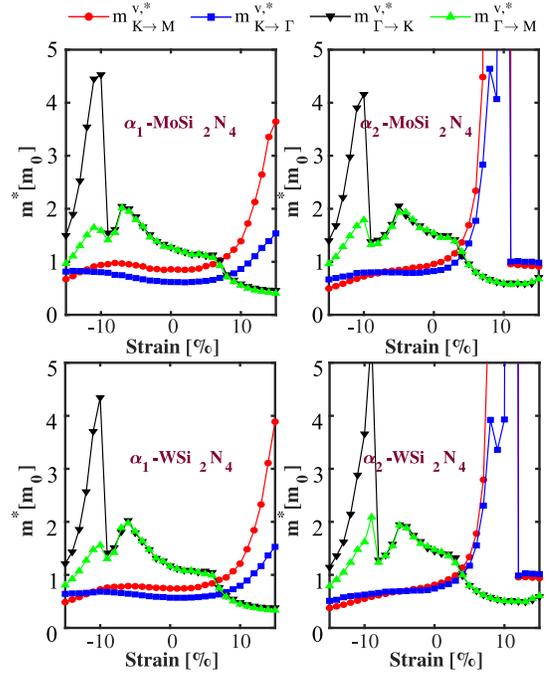}
	\caption{The hole effective masses as a function of strain at two effective valleys (K and $\Gamma$). The effective masses of $\Gamma$-valley along $\Gamma$ to K ($m_{\Gamma\rightarrow K}^{v,*}$) and $\Gamma$ to M ($m_{\Gamma\rightarrow M}^{v,*}$) and effective masses of K-valley along K to $\Gamma$ ($m_{K\rightarrow\Gamma}^{v,*}$) and K to M ($m_{K\rightarrow M}^{v,*}$) are plotted. }
	\label{fig:mass}
\end{figure}

The hole effective mass has been explored under various strains. The effective masses of these two valleys versus strain are reported in Fig. \ref{fig:mass}. Two K- and $\Gamma$-valleys, contribute to the VBM in the most range of strain. The K-valley exhibits the lowest effective mass at compressive strains and one can find from Fig. \ref{fig:energy} that the VBM is located at K-valley in this strain range. The values of the K-valley effective masses increase as the strain changes from compressive to the tensile regime. In addition, in $\alpha_1$ structures two different paths at K-valley demonstrate different effective masses, whereas these two paths possess the same value for $\alpha_2$ ones. The difference of two effective masses in $\alpha_1$ structures increases with a rise in the strain and K-valley shows more anisotropy for tensile strains. At the same time, the effective mass of $\Gamma$-valley decreases with the strain and becomes the lowest effective mass at tensile strains. At the tensile regime, the VBM is located at $\Gamma$-valley. The $\Gamma$-valley demonstrates isotropic effective mass for a wide range of strain and only shows anisotropic effective mass at large compressive strain, where it does not contribute to the VBM. The effective masses for both MoSi$_2$N$_4$ and WSi$_2$N$_4$ behave similarly. However, WSi$_2$N$_4$ demonstrates a little lower effective mass.

\begin{figure*}
	\centering
	\includegraphics[width=0.85\linewidth]{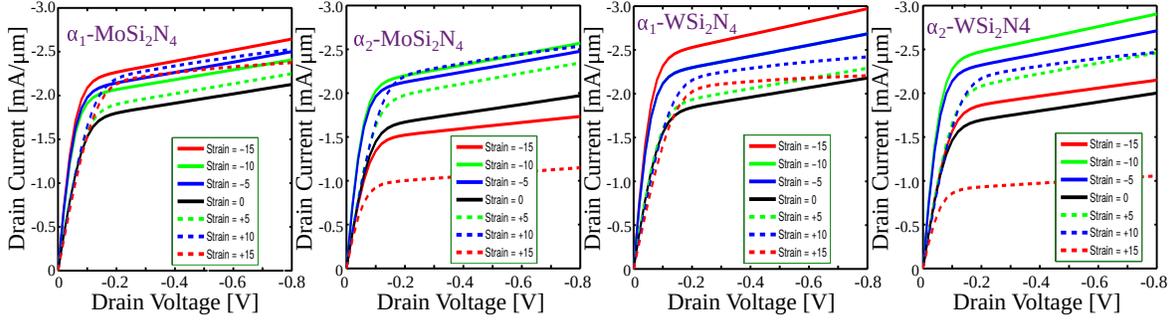}
	\caption{The current-voltage charachteristic of DGFET with MoSi$_2$N$_4$ and WSi$_2$N$_4$ as the channel material under biaxial strain. }
	\label{fig:fig-id-vd}
\end{figure*}

\begin{figure*}
	\centering
	\includegraphics[width=1.0\linewidth]{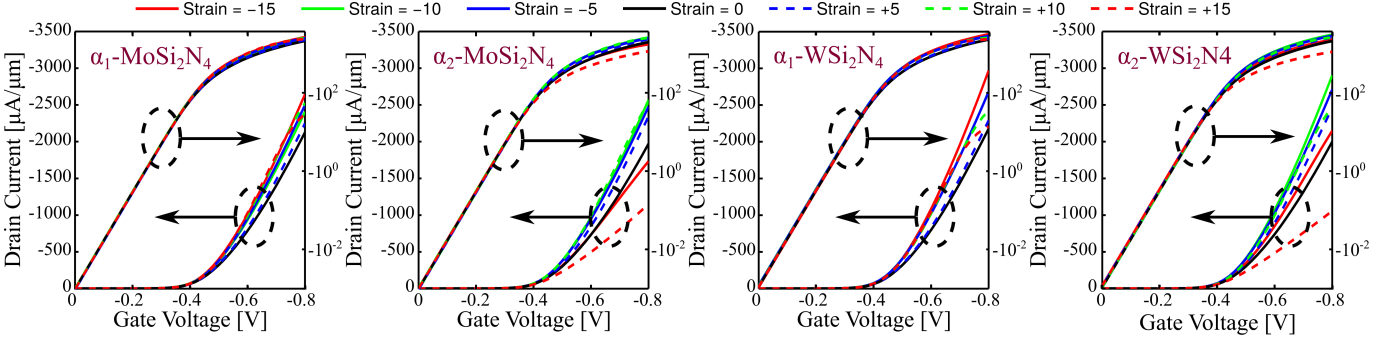}
	\caption{The drain current versus gate voltage of studied DGFET under different values of strains in two linear and logarithmic scales.}
	\label{fig:fig-idvglinear}
\end{figure*}

The electronic properties of the DGFET with strained MoSi$_2$N$_4$ and WSi$_2$N$_4$ as the channel material is investigated through the top of the barrier model \cite{touski2020electrical,manouchehr2020investigation}. These materials indicate p-type behavior at experiment results and here, a p-type FET (PMOS) is also investigated. Drain current ($I_{DS}$) versus drain-source voltage ($V_{DS}$) is drawn in Fig. \ref{fig:fig-id-vd}. $I_{DS}$ increases at small drain voltages and it saturates around $V_{DS}=0.1V$. Then, $I_{DS}$ continues to increase with a small slope. Drain current is also investigated under different values of strains. The equilibrium condition and large compressive strains demonstrate the lowest and highest current in the  $\alpha_1$-structures, respectively. On the other hand, the large tensile and the compressive strain demonstrate the lowest and highest current for the $\alpha_2$-structures, respectively. The equal energy of K- and $\Gamma$-valleys at equilibrium state for all structures results in the high effective mass and small current.

The drain current as a function of gate voltage for different strain values is shown in Fig. \ref{fig:fig-idvglinear}. For better understanding, the drain current is studied at both linear and logarithmic scales. The linear scale clarifies the ON-current performance and follows the I$_{DS}$-V$_{DS}$ behavior. The ON-current, OFF-current and I$\mathrm{_{ON}}$/I$\mathrm{_{OFF}}$ ratio for four compounds are obtained in the range of 2000-2200$\mu A/\mu m$, $10^{-3} \mu A/\mu m$, and 2.0-2.2$\times 10^6$, respectively. These values are comparable with FET based on the other monolayers\cite{sengupta2013performance,yin2015scaling,mukhopadhyay2021effect,marin2018first,ouyang2007electron}. The steep rising of the current in the sub-threshold region is obvious from the logarithmic scale and results in a small sub-threshold swing (SS). The four compounds indicate similar SS in the range of 96-98 mV/dec. Indeed, the SS is controlled by the FET structure more than channel material.

\begin{figure}
	\centering
	\includegraphics[width=0.7\linewidth]{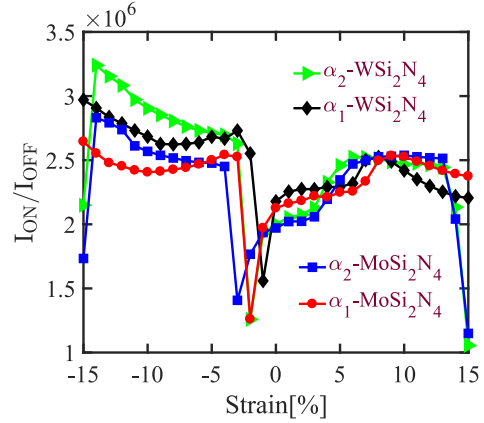}
	\caption{The I$\mathrm{_{ON}}$/I$\mathrm{_{OFF}}$ ratio as a function of strain for DGFETs with studied compounds as the channel material.}
	\label{fig:ss}
\end{figure}

One of the most important parameters of the FETs is I$\mathrm{_{ON}}$/I$\mathrm{_{OFF}}$ ratio. The I$\mathrm{_{ON}}$/I$\mathrm{_{OFF}}$ ratio is over $10^6$ for nowadays technologies. The obtained I$\mathrm{_{ON}}$/I$\mathrm{_{OFF}}$ ratio versus biaxial strain for all four compounds are plotted in Fig. \ref{fig:ss}. All curves demonstrate a minimum at small compressive strains close to the equilibrium condition. At these strains, both K- and $\Gamma$-valley possess the same energy that results in a high effective mass. This high effective mass leads to a small ON-current and consequently small I$\mathrm{_{ON}}$/I$\mathrm{_{OFF}}$ ratio. In addition, the $\alpha_2$-structures display other minimums at -15 and +15$\%$ strains. In these strains, three valleys ($\Gamma$, K and M) have the same energy that results in a high effective mass and low ON-current. However, K- and $\Gamma$-valley are dominant at these strain values. The I$\mathrm{_{ON}}$/I$\mathrm{_{OFF}}$ ratio follows the effective mass behavior out of these minimums. $\Gamma$-valley is dominant over tensile strains and controls the current. The reduction of the effective mass of $\Gamma$-valley at tensile strains increases the ON-current and I$\mathrm{_{ON}}$/I$\mathrm{_{OFF}}$ ratio. One can observe a jump in the I$\mathrm{_{ON}}$/I$\mathrm{_{OFF}}$ ratio for moderate tensile strains. Such a jumping has been seen in the effective mass of $\Gamma$-valley at tensile strains.  On the other hand, the VBM is determined by K-valley at compressive strains. The effective mass of K-valley increases for small compressive strain and decreases for larger compressive strains. I$\mathrm{_{ON}}$/I$\mathrm{_{OFF}}$ ration also decreases for small compressive strains but increases for larger compressive strains.

\section{Conclusion}

The electronic properties and performance of DGFETs based on two different configurations of monolayer MSi$_2$N$_4$ (M = Mo, W) under biaxial strain are investigated. These materials exhibit indirect bandgap with CBM and VBM at K- and $\Gamma$-valleys, respectively. The $\alpha_2$-configurations demonstrate a higher bandgap along with larger effective masses. The $\Gamma$-valley demonstrates equal effective masses along different directions, whereas K-valley has anisotropic effective masses. The band gaps and effective masses can be tuned by in-plane biaxial strain. All compounds show a maximum value of bandgap at compressive strains between -2$\%$ and -4$\%$. Afterward, the bandgap decreases for both compressive and tensile strains. Finally, the current-voltage characteristic of p-type FET with MSi$_2$N$_4$ as the channel has been explored. $I_{DS}$ increases at small values of $V_{DS}$ until its saturation around $V_{DS}=0.1V$. The I$\mathrm{_{ON}}$/I$\mathrm{_{OFF}}$ ratio is larger than $\mathrm{10^6}$ and sub-threshold swing is obtained in the range of 96-98 mV/dec. Furthermore, bi-axial strain can effectively modulate the current-voltage characteristic of DGFETS. The  
I$\mathrm{_{ON}}$/I$\mathrm{_{OFF}}$ ratio with respect to strain demonstrates a minimum at small compressive strains close to equilibrium condition for all materials. $\Gamma$-valley is dominant over tensile strains and controls the current. The reduction of the effective mass of $\Gamma$-valley at tensile strains increases the ON-current and I$\mathrm{_{ON}}$/I$\mathrm{_{OFF}}$ ratio.

%%%%%%%%%%%%%%%%%%%%%%%%%%%%%%%%%%%%%%%%%%%%%%%%%%%%%%%%%%%%%%%%%%%%%%%%%%%%%%%%%%%%%%%%%%%%%%%%%%%%%
%%%%%%%%%%%%%%%%%%%%%%%%%%%%%%%%%%       BIBLIOGRAPHY      %%%%%%%%%%%%%%%%%%%%%%%%%%%%%%%%%%%%%%%%%%
%%%%%%%%%%%%%%%%%%%%%%%%%%%%%%%%%%%%%%%%%%%%%%%%%%%%%%%%%%%%%%%%%%%%%%%%%%%%%%%%%%%%%%%%%%%%%%%%%%%%%
%\bibliographystyle{IEEEtran}
%%\bibliography{FET,TFET,sige,/home/shoeib/BibTex/acronym}
%\bibliography{acronym,DFT,MoS2,sb}


\end{document}